# COMPARISON AND EVALUATION OF DIGITAL SIGNATURE SCHEMES EMPLOYED IN NDN NETWORK

Al Imem Ali[1]

[1] PRINCE ISITC, H. Sousse University of Sousse, 4011 Hammam Sousse, Tunisia

*ABSTRACT*

*It is well known that Named Data networking ensure data integrity so that every important data has to be signed by its owner in order to send it safely inside the network. Similarly, in NDN we have to assure that none could open the data except authorized users. Since only the endpoints have the right to sign the data or check its validity during the verification process , we have considered that the data could be requested from various types of devices used by different people, these devices could be anything like a smartphone, PC, sensor node etc.b, with a different CPU descriptions, parameters, and memory sizes, however their ability to check the high traffic of a data during the key generation and/or verification period is definitely a hard task and it could exhaust the systems with low computational resources. RSA and ECDSA as digital signature algorithms have proven their efficiency against cyber-attacks, they are characterized by their speed to encrypt and decrypt data, in addition to their competence at checking the data integrity. The main purpose of our research was to find the optimal algorithm that avoids the system's overhead and offers the best time during the signature scheme*

.*KEYWORDS*

*RSA, ECDSA, NDN, MSS, Digital signature algorithms, Security,Encryption, Decryption, Merkle scheme, Eliptic Curve*

## 1.INTRODUCTION

The digital signature algorithms uses some sort of complex operations aims to prevent the data from being accessible only for the authorized users, and computing such operations could exhaust the systems with limited computational resources; the problem statement affects the integrity of the data directly and minimizes the security level to facilitate the process of penetration, then the algorithms must be selected depending on the degree to maintain the integrity of the information regardless of the type of device.

The NDN project team [13] considered security as an important factor; the researchers designed a new security model and implemented it not only into the architecture (e.g. communication channel) but also inside the packet that carry the information itself.

Security are implemented directly by using a digital signature in each NDN data packets, and then it's important for us to check the efficiency of these implemented protocols. The data packet could carry any type of information: credit cards number, personal information, and passwords, top secret information etc., then the role of the digital signature is to keep the data packet away from the cyber-attacks.

This new concept will bring back user's trust. There are numerous digital signatures algorithms used in NDN such as RSA and ECDSA characterized by their high level of security and their speed to encrypt and decrypt data according to [1], moreover their efficiency to generate signatures and verify the data integrity with reduced key sizes. Besides RSA and ECDSA, there is also MSS (Merkle signature scheme) defined by [10] which is an interesting alternative for well-established signature schemes such as RSA, and ECDSA proved their eligibility against cyber-attacks e.g. timing attacks.

The remaining part of the paper is organized as follows: Section 1 involves the NDN security concept with details and enumerates the advantages and disadvantages of RSA and ECDSA algorithms besides their limitation. Section 2 present a modular reduction used for accelerating one of those protocols RSA or ECDSA. Section 3 describes the simulation process used to clarify and illustrate the differences between RSA and ECDSA. The paper is concluded in section 4.

## 2. DIGITAL SIGNATURE

The digital signature provides a means of integrity checking [11]. This is done to provide assurance for the receiver that the data was in fact sent by the assumed party. The integrity plays a critical role in virtual society and it's important to protect it from coming out to the public ensure data integrity so that every important data has to be signed by its owner in order to send it safely inside the network. Similarly, NDN assure that none could open the data except authorized users. The digital signatures in use today can be classified according to the high underlying mathematical problem, which provides the basis for their security:

 • Integer Factorization (IF) problem: RSA signature schemes can be seen as an example under this classification[6].
 • Elliptic Curve (EC) discrete logarithm problem represented in the Elliptic curve digital signature algorithm[4].

RSA and ECDSA as digital signature algorithms have proven their efficiency against cyber-attacks, they are characterized by their speed to encrypt and decrypt data, in addition to their competence at checking the data integrity

### 2.1. RSA

"The RSA algorithm was developed at Massachusetts Institute of Technology (MIT) in 1977 by Ron Rivest, Adi Shamir and Leonard Adelman[6] ". The RSA concept is based on the factorization of big numbers which means the larger sequence of numbers you have, the more you are protected. The RSA provides a strong security; therefore an adversary should not be able to break RSA by factoring due to its complexity and large keys. RSA is used to encrypt/decrypt data and also has the ability to sign and/or verify the data packets. RSA does not mandate the use of a particular hash function, so the security of the signature and encryption are partly dependent on the choice of hash function used to compute the signature.

### 2.2. ECDSA

"The Elliptic Curve Digital Signature Algorithm (ECDSA) is the elliptic curve analogue of the Digital Signature Algorithm (DSA). It was accepted in 1999 as an ANSI standard, and was accepted in 2000 as IEEE and NIST standards. It was also accepted in 1998 as an ISO standard, and is under consideration for inclusion in some other ISO standards, the range of parameters offered by the standard can provide security for a number of years, provided the lowest figures are gradually discarded, taking into account the progress of computing power" [14].

Unlike the ordinary discrete logarithm problem and the integer factorization problem, no sub exponential-time algorithm is known for the elliptic curve discrete logarithm problem. Elliptic curve digital signature (ECDSA) was developed in 1985 by Neal Koblitz and Victor Miller.
"ECDSA schemes provide the same functionality as RSA schemes including sign and/or verify signed packets. There are some environments where 1024-bit RSA cannot be implemented, while 192-bit ECDSA can. For this reason, the strength-per-key-bit is substantially greater in an algorithm that uses elliptic curves" [4].

The Elliptic curve cryptography (ECC) is a very efficient technology to realize public key cryptosystems and public key infrastructures (PKI). The security of a public key system using elliptic curves is based on the difficulty of computing discrete logarithms in the group of points on an elliptic curve defined over a finite field according to the [7]. The claim is that a 192 bit ECDSA key is similar to a 1024 bit RSA key in terms of the security that it offers.

The performance tests therefore make comparisons according to these claims and attempt to provide more insight into the most suitable public key cryptography algorithm for a mobile framework given its limitations.

The following table from the article wrote by [12], represents the recommended comparative key lengths for RSA and ECDSA used in software implementations. RSA needs larger key length but instead ECDSA requires significantly smaller key size with same level of security which offers faster computations and less storage space. ECDSA ideal for constrained environments: Tablets, Smart phones, RFID systems, Sensors etc.

Table 1. Equivalent key lengths for RSA and ECDSA

| RSA key length (bits) | ECDSA key length (bits) |
|---|---|
| 1024 | 192 |
| 2048 | 256 |

## 2.3. RSA and ECDSA digital signature schemes

The essential elements of the digital signature schemes for RSA and ECDSA which defined by three computational procedures or algorithms:

### 2.3.1. Key generation procedure

The key generation procedure is used to generate the keys that are used by the signing procedure and the verifying procedure. Each time it is used the procedure generates a key pair consisting of a Private/signature key and the corresponding Public/verification key. It is important to note that the key generation procedure uses a random number generator and will generate a different pair each time it is used. SK is always known as the secret key because in applications the signing key is kept secret. PK is always known as the public key, the verification key is distributed to all users who want to verify signatures.

**2.3.2. Signing procedure**

The producer generates a signing process to transform/change the data from its original format to a new protected form. Each time it is used the procedure takes as input a signature key generated using the key generation procedure and data from some pre-determined data space. The signing procedure transforms the data and produces a signature as an output for the producer or the legal owner.

**2.3.3. Verifying procedure**

Consumers who receive desired data packet in reverse need to be able to check that the signature appended to the message is correct, in the sense that it is a value which would be produced if the signing procedure was applied to the received data packet using the Producer's signing key. The verifying procedure takes as input the data and signature together with the public key of the purported consumer and then either accepts or rejects the signature. If the verifying procedure outputs 'Accept,' then the message are accepted as valid; otherwise it is rejected as invalid and the consumer sends a new interest packet.

## 2.4. Comparison of ECDSA with RSA

Here are some remarkable differences between RSA and ECDSA investigated by Khalique in[2]:

- ECDSA offers same level of security with smaller key sizes.
- Data size for RSA is smaller than ECDSA.
- Encrypted message is a function of key size and data size for both RSA and ECDSA. ECDSA key size is relatively smaller than RSA key size, thus encrypted message in ECDSA is smaller.
- Computational power is smaller for ECDSA.
- ECDSA provides faster computations and less storage space
- ECDSA key sizes are so much shorter than comparable RSA keys
- The length of the public and private keys is much shorter in ECDSA. This results in faster processing times, and lower demands on memory and bandwidth.

Note: "Some researchers have found that ECDSA is faster than RSA for signing and decryption process, however ECDSA is a bit slower for signature verification and encryption"[8].

## 2.5. Advantages of ECDSA

The ECDSA offered remarkable advantages over other cryptographic system mentioned by [2].

- It provides greater security with smaller key sizes.
- It provides effective and compact implementations for cryptographic operations requiring smaller chips.
- Due to smaller chips less heat generation and less power consumption.
- It is mostly suitable for machines having low bandwidth, low computing power, less memory.
- It has easier hardware implementations.

**2.6. RSA and ECDSA limitations**

Here are the most RSA and ECDSA remarkable limitations stated in [2]:

- Key generation is very slow.
- Speed of encrypting of data is slow.
- Message length should be less than the bit length otherwise algorithm will fail.
- RSA is factorization based algorithm so that every time RSA initialization takes two large prime number p and q.

## 3.THE MONTGOMERY REDUCTION

"Peter Montgomery has devised a way to speed up arithmetic in a context in which a single modulus is used for a long-running computation"[5]. Montgomery multiplication algorithm reduces the computation time taken by a computer when there are a large number of multiplications to be [9]. The Montgomery reduction algorithm above uses the shift and/or adds operation to change the mode of calculation and makes it easier. The method can be used to reduce memory consumption, the execution time. The Montgomery method uses modular reduction to minimize the cycles taken by the signature schemes.

## 4.EXPERIMENTAL RESULTS

Our objective was to compare two signature algorithms RSA and ECDSA used in NDN, using an (Intel core i5-2450 M CPU 2.50GHz and 4GB -RAM) machine. We combined NDNx simulator with Open SSl library in order to measure:

- The required time to sign the data packet.
- The required time to verify the data packet.
- The required time to generate the private and/or public keys.

Moreover, we used the Montgomery simulator to compare the ECDSA computation tasks by calculating the necessary operations and/or cycles between the ordinary and the Montgomery multiplication methods in order to resample the algorithm and speed up ECDSA.

During the simulation process we observed that RSA performs better with shorter keys while ECDSA showed their slowness during the verification process. This weakness point needed to be solved, thus we implemented the montgomery multiplication technique in order to speed up ECDSA and upgrade its performance.

**4.1. RSA speed Performance**

The obtained results showed that the verification process do more faster than the signing process, moreover we noticed that the more we increase the key size, there were no big difference during the verification process, the obtained results were too close to each other for the different key sizes which means that even if the RSA key size got doubled RSA still fast compared with the signing process.

The RSA-512 and RSA-1024 recorded the best time during the verification and/or signing process compared with RSA-2048 and RSA-4096. We've got unreasonable that won't suit the specifications of embedded systems.

Table 2. RSA speed performance time statics

| Key length (bits) | RSA-512 | RSA-1024 | RSA-2048 | RSA-4096 |
|---|---|---|---|---|
| Sign per key size(sec) | 0.00006 | 0.000205 | 0.001508 | 0.010717 |
| Sign per key size(sec) | 0.000005 | 0.000013 | 0.000045 | 0.000175 |

### 4.2. RSA key generation

The following table represents how much time taken by the algorithm to generate the Public/Private keys using different key lengths. The mean time generated by RSA-512 and RSA-1024 recorded the fastest time due to its small key sizes compared with RSA-2048 and RSA-4096.

The RSA-4096 key generation was out of control and it exhausted the machine performance during the simulation process and increased the system's overhead, and registered the slowest time during the signature process, wich could exhaust the battery lifetime and energy on wireless devices; Generating the RSA keys to encrypt the data by the producers especially if some of them are using a wireless device will lockup the system for some time and may completley drain the battery.

Table 3. RSA key generation time statics

| Key Length(bits) | RSA-512 | RSA-1024 | RSA-2048 | RSA-4096 |
|---|---|---|---|---|
| Key generator (sec) | 0.132 | 0.431 | 2.559 | 112.3 |

### 4.3. ECDSA speed performance

During the ECDSA speed performance, we observed that the signature generation and/or verification time does not differ until the larger key sizes: ECDSA-256→ECDSA-521. The following table represents the generated time for each key size. We observed that ECDSA overpass RSA during the signature generation process. However RSA defeat ECDSA in performance during the verification process. ECDSA-192 and ECDSA-160 registered the best time during the sign/verification process. The ECDSA keys keeps the same level of security as RSA and provides quicker computation, lower power consumption, memory and bandwidth savings as an addition.

Table 4. ECDSA speed performance time statics

| Key length(Bits) | ECDSA-160 | ECDSA-192 | ECDSA-224 | ECDSA-265 | ECDSA-384 | ECDSA-521 |
|---|---|---|---|---|---|---|
| Sign per key size(sec) | 0.0003 | 0.0003 | 0.0007 | 0.0009 | 0.0016 | 0.0033 |
| Verify per key size(sec) | 0.0015 | 0.002 | 0.0024 | 0.0039 | 0.0082 | 0.018 |

**4.4. ECDSA key generation**

ECDSA can create keys in superior speed to RSA comparable key lengths. The overall comparison showed that generating a 2048 bit key for RSA takes significantly longer than generating a 1024 bit key.

Table 5.  ECDSA key generation time statics

| Key length(Bits) | ECDSA-160 | ECDSA-192 | ECDSA-224 | ECDSA-265 | ECDSA-384 | ECDSA-521 |
| --- | --- | --- | --- | --- | --- | --- |
| Key generation(sec) | 0.161 | 0.165 | 0.229 | 0.305 | 0.799 | 1.584 |

However, This is not the case for ECDSA where there is small increase in the execution time. This is already expected since there is a 1024 bit difference between the two RSA keys and only small difference between the two ECDSA keys.

**4.5. The Montgomery reduction**

The Montgomery reduction algorithm are applied to speed up modular exponentiation and use shift operations in the place of modular reductions. One such algorithm is the square and multiply algorithm from [3]:

    1. Set result = 1. If e = 0 then return (result).
    2. Set A = m.wa
    3. If k0 = 1 then set result = m
    4. For i from 1 to t do the following
        4.1    Set A = A² mod n
        4.2    If ei = 1 then set result = A. result mod n.
    5. Return ← result

The standard multiplication method is a completely hard task to accomplish its work, however the Montgomery method takes shorter path to get the intended results as the same as the standard method.

The algorithm above are up to the same results from each side. Our interests were in replacing the current ECDSA algorithm with a modified algorithm which we called "modified ECDSA" and to show the differences between.

We suspended RSA algorithm from upgrading its performance for the reason that installing such algorithm on light-weight devices will adversely affect their performance and delay the decryption process. ECDSA in counterpart could be a replacement for RSA system, their comptability to be installed in any system with different memory sizes and CPU description and parameters, ECDSA provide the same level of security as RSA but with shorter keys: The smaller key sizes of ECDSA potentially allow for less computationally able light-weight devices and wireless systems to use cryptography for secure data transmissions, data verification and offers less heat generation and less power consumption, less storage space and offers an optimized memory and bandwidth and faster signature generation. To reduce the cycles used by the ECDSA we used this algorithm from [3]:

International Journal of Embedded systems and Applications(IJESA) Vol.5, No.2, June 2015

    1. M = Mont (m, R2 mod n), result = R mod n where Mont (u, v) is uvR-1 mod n
    2. For i from t down to 0 do the following:
       2.1 result = Mont (result, result)
       2.2 If $e_a$. = 1 then A = Mont (result, M)
    3. result = Mont(result,1)
    4. Return(result).

The Montgomery reduction algorithm above uses the shift and/or adds operation to change the mode of calculation and makes it easier. The method can be used to reduce memory consumption, the execution time.

The Montgomery method uses modular reduction to minimize the cycles used by each process: signing/verification process. We've made four scenarios and for each scenario we've used a different bit lengths sequence. Since it was hard to calculate that huge number of exponents our research were only limited to 64 bit lengths.

We've been through calculating the operations needed per each method, the Montgomery method uses an extra module called Add/shift multiplier used to optimize the big charge of operations by the standard multiplication method.

**4.5.1. Scenario 1: Standard method vs. Montgomery method statics: 16 bits length**

The first scenario showed that the Montgomery method took fewer cycles than the standard method, the Add and/or shift module optimized the modular reduction to only 2 % (see figure 1) from the global operations.

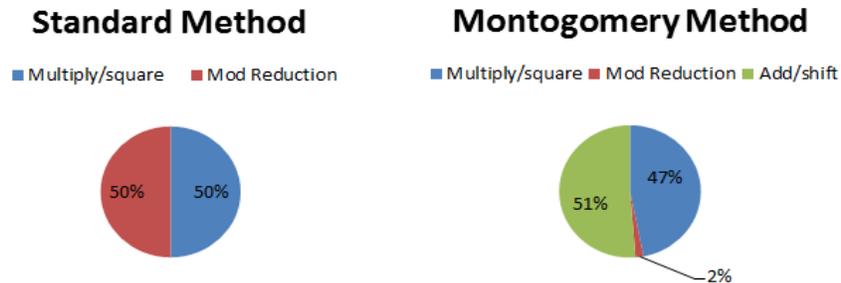

Figure1. Standard /Montgomery Method calculation Percent: 16 bits

In addition it saved 42 cycles against the standard method. The number of cycles saved, aims to reduce the CPU execution time and to speed up the digital signatures procedure such as verification or generation time.

International Journal of Embedded systems and Applications(IJESA) Vol.5, No.2, June 2015

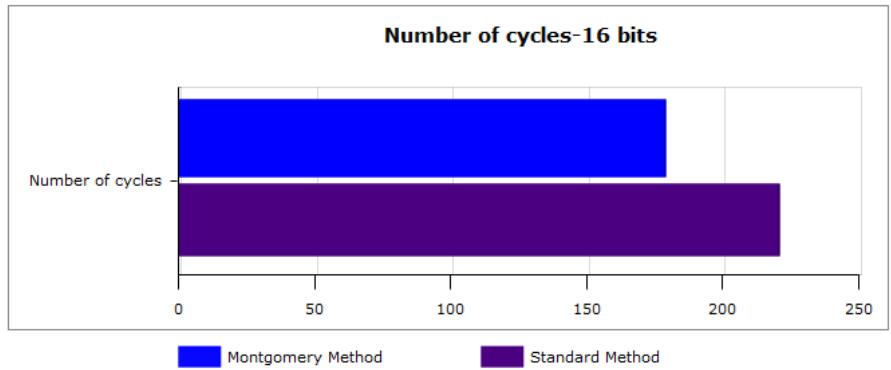

Figure 2. Standard method vs. Montgomery method statics: 16 bits length

**4.5.2. Scenario 2: Standard method vs. Montgomery method statics: 32 bits length**

The second scenario showed a reduction by 121 cycles, the Add and/or shift module gain another 79 compared with the last Montgomery results. The number of the modular reductions used by the standard method are greater than the Montgomery's, then it delays the computation tasks.

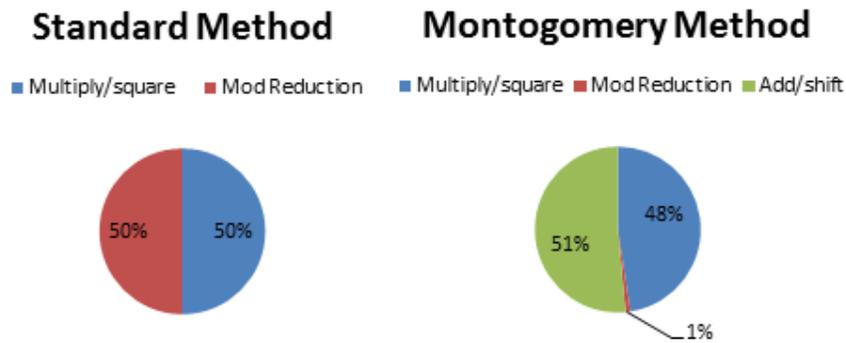

Figure 3. Standard /Montgomery Method calculation Percent: 32 bits

The second scenario showed a reduction by 121 cycles, the Add and/or shift module gain another 79 compared with the last Montgomery results. The number of the modular reductions used by the standard method are greater than the Montgomery's, then it delays the computation tasks.

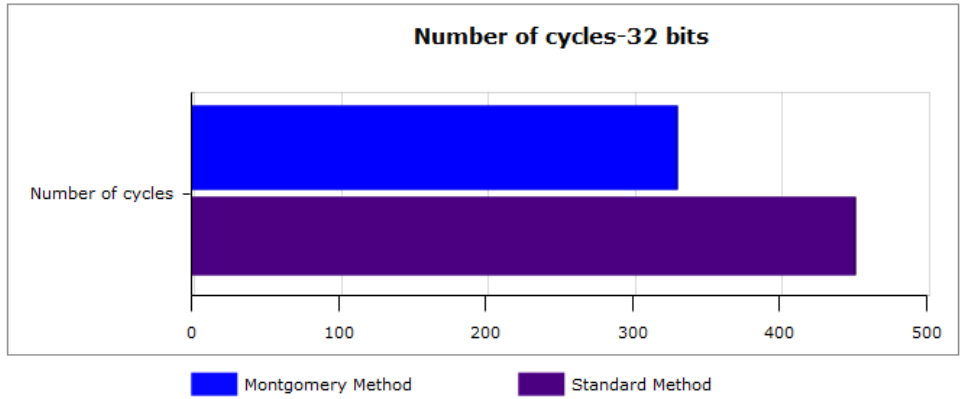

Figure 4. Standard method vs. Montgomery method statics: 32 bits length

### 4.5.3. Scenario 3: Standard method vs. Montgomery method statics: 48 bits length

The third scenario is a complementary for the other two scenarios; we kept getting good results since the Montgomery method succeeded to reduce the number of cycles. This time we got rid of 188 cycles. Moreover the modular reduction used by the standard Method is greater than the Montgomery's, and then it delays the computation tasks.

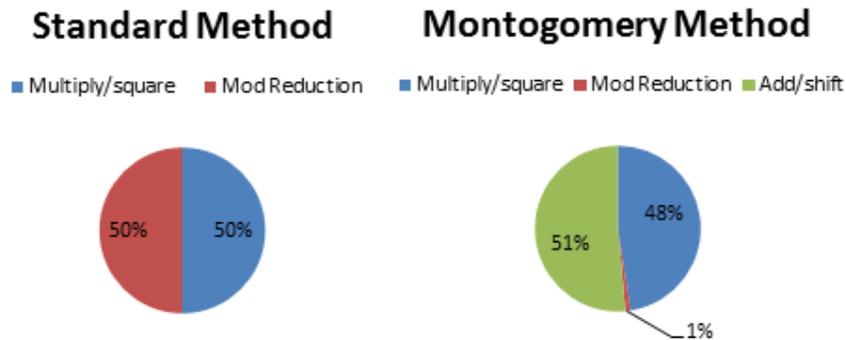

Figure 5. Standard /Montgomery Method calculation Percent: 48 bits

The shift and/or add operations might be greater, but it helps to shorten the numbers of modular reduction. The numbers of modular reduction used in the standard method are extremely hard task. The shift and/or add method aims to reduce the numbers of operations used in modular reduction.

International Journal of Embedded systems and Applications(IJESA) Vol.5, No.2, June 2015

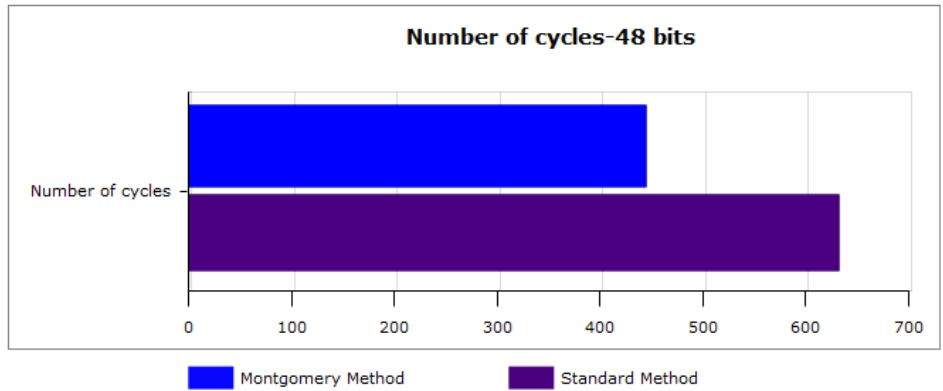

Figure 6. Standard method vs. Montgomery method statics: 48 bits length

**4.5.4.Scenario 4: Standard method vs. Montgomery method statics: 64 bits length**

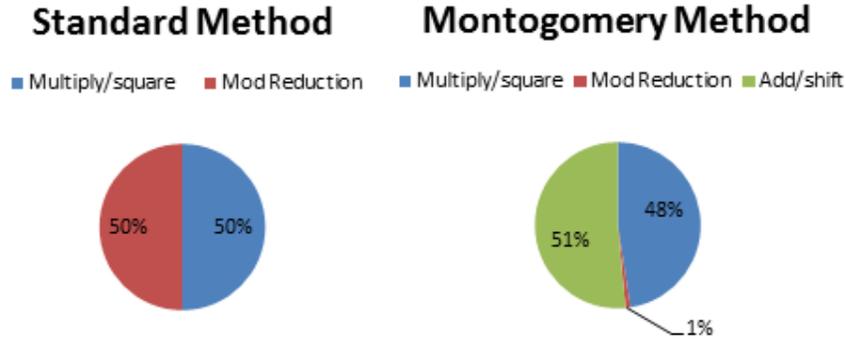

Figure 7. Standard /Montgomery Method calculation Percent: 64 bits

The last scenario showed big differences from the previous scenarios we got rid of 297 cycles which means that we took the right road to prove that the Montgomery could be a replacement of the standard multiplication method.

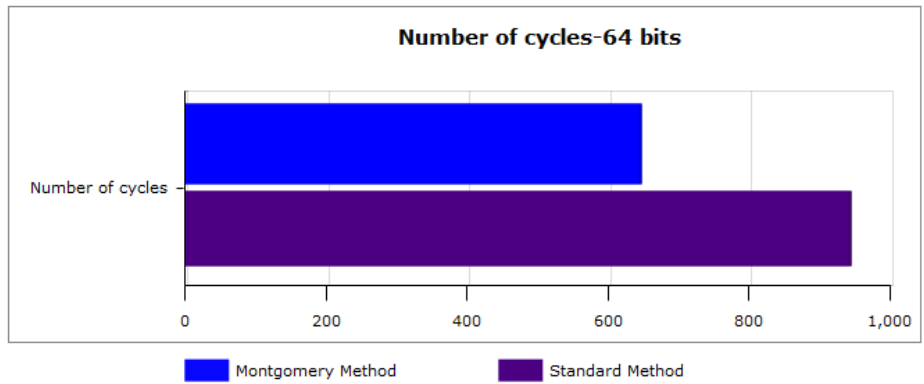

Figure 8. Standard method vs. Montgomery method statics: 64 bits length

One of our objective were to equilibrate ECDSA and RSA and improve ECDSA recorded time. The Montgomery method offers an optimized multiplication sequences which aims to speed up the regular ECDSA algorithm process. the next part shows that the Montgomery method is an effective way to accelerate the regular ECDSA.

### 4.6 Modified ECDSA speed Performance

The results are shown in Table 6 represents the Modified ECDSA speed performance during the verification and signing process. We've got reasonable results for both sides as we can see 0.0001 ms for signing and between 0.0003 ms and 0.0002 ms for verification. It is safe to assume that embedded systems with low computational ressources will be more than capable of verifying ECDSA digital signatures. In terms of speed our verification routine outperforms RSA and ECDSA for all common security parameters.

Table 6. Modified ECDSA speed performance time statics

| Key length(Bits) | ECDSA-160 | ECDSA-192 | ECDSA-224 | ECDSA-265 | ECDSA-384 | ECDSA-521 |
|---|---|---|---|---|---|---|
| **Sign per key size(sec)** | 0.0001 | 0.0001 | 0.0001 | 0.0001 | 0.0002 | 0.0005 |
| **Verify per key size(sec)** | 0.0002 | 0.0003 | 0.0003 | 0.0005 | 0.0007 | 0.001 |

## 5. DISCUSSION

### 5.1 Performance Comparison: Standard Method, Montgomery Method

The Montgomery multiplication method showed interesting results and succeeded to reduce the number of operations with the Add/shift method. The number of the modular reduction used by the standard method is greater than the Montgomery's, and then it delays the computation tasks. The shift and/or add operations might be greater, but it helps to shorten the numbers of modular reduction.

The numbers of modular reduction used in the standard method are extremely hard task then the shift and/or add method aims to reduce the numbers of operations used in modular reduction and so it could accelerate the digital signature process such as ECDSA and improve its performance and release the system from additional tasks. The first table summarizes the Montgomery vs. Standard method performance:

Table 7. Performance Comparison: Standard Method, Montgomery method

| Method/Factors | Number of operations | Cycles |
|---|---|---|
| **Standard Method** | Less | Many |
| **Montgomery Method** | Greater | Optimized |

## 5.2 Performance Comparison: RSA, ECDSA, MODIFIED ECDSA

The results that were obtained for all the performance measurements have been categorised according to the dependent variables. The goal with this round of tests was to provide recommendations regarding the chosen algorithms with respect to their performance and compared to the level of security provided.

The second table summarizes the digital signatures algorithms including the recommended domains:

Table 8. Performance Comparison: RSA, ECDSA, Modified ECDSA

|  | Security | Complexity | Domain | Key Creator | Execution Time | Verify/s | Sign/s |
|---|---|---|---|---|---|---|---|
| **RSA** | High | Integer Factorization | PC, laptops, Super computers | Fast | Slow | Fast | Fast |
| **ECDSA** | High | Discrete logarithm | Light-weight devices | Faster | Fast | **Slow** | **Fast** |
| **Modified ECDSA (Our)** | High | Discrete logarithm | Light-weigh devices | Faster | Fast | **Fast** | **Faster** |

We suspended RSA algorithm from upgrading its performance for the reason that installing such algorithm on light-weight devices will adversely affect their performance and delay the decryption process. ECDSA in counterpart could be a replacement for RSA system, their comptability to be installed in any system with different memory sizes and CPU description and parameters, ECDSA provide the same level of security as RSA but with shorter keys: The smaller key sizes of ECDSA potentially allow for less computationally able light-weight devices and wireless systems to use cryptography for secure data transmissions, data verification and offers less heat generation and less power consumption, less storage space and offers an optimized memory and bandwidth and faster signature generation. One of our objective were to equilibrate ECDSA and RSA and improve their times. The Montgomery method we employed offers an optimized multiplication sequences which aims to speed up the regular ECDSA algorithm process. Our method changed the state of the regular algorithm's verification time from slow to fast and the algorithm's sigining time from fast to faster.

The following discussion profiles the above comparison according to the steps generated by each algorithm.

### 5.2.1. Key generation

The key creation for ECDSA was significantly faster than RSA due to the difference in the key lengths. The comparison shows very small ratios whatever the security level. This means the key generation with ECDSA is always faster than the key generation with RSA. The RSA keys are generated using large prime numbers thus take significantly longer than the smaller ECDSA keys that are generated.

**5.2.2. Execution Time**

In terms of Execution time, we found that the difference between RSA and ECDSA was significant the resulting graph in Table 3 and Table 5 showing the execution time with standard deviation that RSA takes significantly longer compared to ECDSA to generate its key pair. This result is expected since the RSA keys are significantly larger than the ECDSA keys.

**5.2.3. Verification Time**

In terms of verification time we found that the difference between regular ECDSA and RSA was significant to. This is most likely due to the fact that the regular ECDSA uses a complex operations rather than RSA. For RSA the verification period is very fast since it is simpler in terms of cryptographic computations as only a minimum of modular multiplications is necessary. The work on resampling the ECDSA operations would speed up the signatures algorithm while maintaining the same security level, the Montgomery method we implemented succeeded to to accelerate the signature scheme and gave us better performance by minimizing the time required to verify the number of packets of each data and succeeded to reduce the number of operations during the computation tasks and made it more simple. Our algorithm( Modified ECDSA) has shorten the diffrences between ECDSA and RSA in terms of signing and verification time which could lead us to a new level where we can group security, speed, stability and comptabilty together.

**5.2.4 Signature Time**

The tests that were conducted for both RSA and ECDSA have shown that the RSA time signature performs poorly compared with ECDSA and modified ECDSA ones. The time required for RSA operations to generate signatures quickly rises due to its larger keys which is not acceptable and could delay the packets from being transfered. The time required for ECDSA times also rise, but at a much slower rate due to the convergence between the keys size.

## 5.CONCLUSION

RSA and ECDSA used to protect the data packet inside the NDN network and to recommend the preferred one depending on the results we've got. In addition, we considered that the time is an important factor that a user wouldn't wait the whole day waiting for encrypting and decrypting the data, the work on resampling the operations would speed up the signatures algorithm while maintaining the same security level, we present the Montgomery method aims to accelerate the signature scheme for better performance and the reduction of the wasted time. To compare the evaluation performance of the RSA and ECDSA digital signatures we used Open-SSL for comparison, and as results we found that:

The key generation time for ECDSA was significantly faster than RSA due to the difference in the key lengths. The RSA keys that are generated using large prime numbers thus take significantly longer than the smaller ECDSA keys that are generated.

The execution time between RSA and ECDSA was significant. This result is expected since the RSA keys are significantly larger than the ECDSA keys.

The verification time between regular ECDSA and RSA was significant to. This is most likely due to the fact that the regular ECDSA uses a complex operations rather than RSA. Our algorithm( Modified ECDSA) has shorten the diffrences between ECDSA and RSA in terms of signing and verification time which could lead us to a new level where we can group security, speed, stability and comptabilty together.

The time required for RSA operations to generate signatures quickly rises due to its larger keys which is not acceptable and could delay the packets from being transfered. The time required for ECDSA times also rise, but at a much slower rate due to the convergence between the keys size. The Montgomery method uses fewer cycles in order to speed up the execution time. Then the Montgomery method we've integrated reduced the number of operations during the computation task and succeeded to accelerate the ECDSA sign and/or verification process.

As future works, we considered that the NDN network will be fully loaded by signed packets coming from different sources and/or destinations. The verification of each signature is definitely a hard task; we considered that verifying a group of signed packets together aims to reduce the verification time. As a solution we propose 'batch verification' technique in order to put together multiple signatures in the same queue for global verification and time saving. However the batch size is limited to a specific size and a specific number of signatures; in order to solve that we propose to compress each signature in order to add more inside the batch. The batch verification can be integrated inside one of the digital signature protocols such as: ECDSA. Then our idea summarized in verifying the maximum number of signatures instantly.

**Author:**

**Ali Al Imem** is currently a Researcher at Higher Institute of Computer Sciences and Communication Techniques, Hammam Sousse , Tunisia. Mr. Ali received his MSc. in Networks from ISITCOM, Tunisia. Mr Ali's areas of interest include Embedded systems, Cryptography and Cloud Computing.
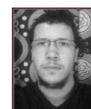